\documentclass{rmf-d}
\usepackage {nopageno, 
rmfbib, multicol,times,epsf,amsmath,amssymb,cite}
\usepackage[latin1]{inputenc}
\usepackage[]{caption2}
\usepackage{graphics}
\usepackage{graphicx}
\usepackage{hyperref}
\usepackage{lipsum}
\usepackage{wrapfig, blindtext}
\usepackage{comment}
\usepackage{tabularx}
\usepackage{float}
\def\rmfcornisa{}

\def\rmfcintilla{}
\clearpage \rmfcaptionstyle 
\newcommand\york{Department of Physics and Astronomy, York University, Toronto, Ontario, M3J 1P3, Canada}
\newcommand\bern{Albert Einstein Center, Universit\"at Bern ,CH-3012 Bern,  Switzerland}
\newcommand\cern{Theory Department, CERN, CH-1211 Geneva, Switzerland}
\newcommand\nycu{Institute of Physics, National Yang Ming Chiao Tung University, 30010 Hsinchu, Taiwan}
\newcommand\kmyork{Department of Mathematics and Statistics, York University, Toronto, Ontario M3J 1P3, Canada}

\newcommand\adelaide{CSSM, University of Adelaide, Adelaide, SA, 5005, Australia}

\begin{document}
\title{   Good and bad diquark properties and spatial correlations in lattice QCD
\vspace{-6pt}}
\author{ A.~Francis$^{a,b,c}$,\; Ph.~de~Forcrand$^b$,\; R.~Lewis$^d$ \;and K.~Maltman$^{e,f}$     }
\address{$^a$\bern,\\ $^b$\cern,\\ $^c$\nycu,\\ $^d$\york,\\ $^e$\kmyork,\\ $^f$\adelaide  }


\maketitle
\begin{abstract}
\vspace{1em} We study good, bad and not-even-bad diquarks on the lattice in a gauge-invariant formalism in full QCD. We establish their spectral masses with short extrapolations to the physical point, observing agreement with phenomenological expectations. We find that only the good diquark has attractive quark-quark spatial correlations, with spherical shape and size $\sim0.6$~fm. Our results provide quantitative support for modelling the low-lying baryon spectrum using good light diquark effective degrees of freedom.   \vspace{1em}
\end{abstract}
\keys{  diquarks, exotic hadrons, lattice QCD \hfill CERN-TH-2021-223 \vspace{-4pt}}
\pacs{   \bf{\textit{11.15.Ha, 12.38.Gc }}    \vspace{-4pt}}

\begin{multicols}{2}

Diquarks as a concept have existed for a long time and actually pre-date QCD \cite{Lichtenberg:1967zz}. In spite of their long history of successes in describing low-lying baryons and exotics, experimental evidence has been difficult to obtain, however.
Formally their interpolating operator may be written as
\begin{equation}
    D_\Gamma =q^c C \Gamma q'
\end{equation}
where $q,q'$ denote two different quark flavors, $c,C$ indicate charge conjugation and $\Gamma$ acts on Dirac space. A list of their possible quantum numbers is given in Table~1.

One particular prediction for them is that light quarks can take on a special, so-called "good", $(\bar 3_F, \bar 3_c, J^P=0^+)$ diquark configuration, see e.g. \cite{Jaffe:2004ph}. In this configuration the two quarks experience a unique attractive effect when compared to other channels. 
This attraction, and associated formation of a diquark substructure in certain hadrons, could in turn explain the pattern of observed states in the low-lying baryon spectrum. The binding energies of recently predicted doubly-heavy tetraquarks \cite{Francis:2016hui,Karliner:2017qjm,Eichten:2017ffp,Czarnecki:2017vco,Francis:2018jyb,Hudspith:2020tdf,Junnarkar:2018twb,Bicudo:2012qt,Bicudo:2015kna,Bicudo:2015vta,Bicudo:2016ooe,Bicudo:2021qxj,Leskovec:2019ioa}, for example, could be partially explained by this effect. Furthermore, whether or not diquark substructures are formed within baryons, especially the nucleon, has been subject to long debate.

Even though diquarks are well founded in QCD, non-perturbative, ab initio input, in particular through lattice simulations, is scarce. The reason is that they are coloured objects, i.e. not gauge-invariant, and the lattice cannot access them directly.
In this contribution, based on the work of \cite{Francis:2021vrr}, we address this issue by forming gauge-invariant probes of diquark properties through embedding them in hadrons that contain in addition a single static quark. The mass of this infinitely heavy quark cancels exactly in mass differences. Additionally, this configuration can be used to define a measure for the diquark structure through density-density correlations. 

\hspace{-4ex}
\begin{minipage}[t!]{0.49\textwidth}
\begin{center}
    \begin{tabular}{cccc}
         \hline\hline
         $J^P$ & C & F & Op: $\Gamma$ \\ \hline
         $0^+$& $\bar 3$&$\bar 3$& $\gamma_5$, $\gamma_0\gamma_5$\\
         $1^+$& $\bar 3$&$6$& $\gamma_i$, $\sigma_{i0}$\\ 
         $0^-$& $\bar 3$&$6$& $1\!\!1$, $\gamma_0$ \\ 
         $1^-$& $\bar 3$&$\bar 3$& $\gamma_i\gamma_5$, $\sigma_{ij}$\\
         \hline\hline
    \end{tabular}
    \end{center}
    {Table 1. {\it Diquark operators and quantum numbers. The first row denotes the so-called "good", the second the "bad" and the final two the "not-even-bad" diquarks.}}
\end{minipage}

\section{Lattice calculation}

Our lattice simulations are performed with $n_f=2+1$ dynamical quark ensembles. The light sea quarks are an isospin doublet, denoted $\ell=u=d$. The strange quark, $s$, mass is held fixed near its physical value while the light quark mass is varied. The corresponding pion mass values are
 $m_\pi=164,~299,~415,~575$ and $707$~MeV. The lattice size in all cases is $L^3 \times T = 32^3\times 64$ with a lattice spacing of $a=0.090$~fm. The dynamical ensembles were generated by 
PACS-CS~\cite{Aoki:2008sm,Namekawa:2013vu} and are publicly available 
from the JLDG repository~\cite{JLDG}. 
To connect with previous studies \cite{Alexandrou:2005zn,Alexandrou:2006cq} we also generated a set of quenched simulations with coupling $\beta=6.0$ and a valence pion mass 
$m_\pi^v=909\,\rm{MeV}$.

\subsection{Diquark spectroscopy}
One possibility of researching diquarks through lattice simulations is to use gauge-fixed approaches \cite{Hess:1998sd,Bi:2015ifa,Babich:2007ah,Teo:1992zu,Negele:2000uk,Alexandrou:2002nn}. However, in this situation masses and sizes become gauge-dependent quantities, which hampers their use in applications.

\hspace{-4ex}
\begin{minipage}[t!]{0.49\textwidth}
\begin{center}
    \begin{tabular}{ccc}
         \hline\hline
         All in [MeV]& {$\delta E_{\rm{lat}}(m_\pi^{\rm{phys}})~~$} & {$\delta E_{\rm{pheno}}~~$}\\\hline
         $\delta (1^+-~0^+)_{ud\,}$ & 198(4) & 206(4)\\
         $\delta (1^+-~0^+)_{\ell s \,}$ & 145(5)& 145(3)\\
         $\delta (1^+-~0^+)_{ss'}$ & 118(2) &\\ \hline
         $\delta(Q[u d]_{0^+} - \bar{Q} u)$ & 319(1)& 306(7)\\
         $\delta(Q[\ell s]_{0^+} - \bar{Q} s)$ & 385(9)& 397(1)\\
         $\delta(Q[\ell s]_{0^+} - \bar{Q} \ell)$ & 450(6)& \\
         \hline\hline
    \end{tabular}
    \end{center}
    {Table 2. {\it Diquark-diquark and Diquark-quark mass differences.
    The phenomenological results are derived from \cite{Zyla:2020zbs}. Their errors are estimated via the difference between results including the charm and the bottom quark, respectively, while the central value is given by the bottom quark result.}}
\end{minipage}

\vspace{2ex}

An alternative is to embed the diquark in a baryon containing a static, i.e., infinitely heavy quark, 
Q, leading to the Euclidean-time-dependent correlator:
\begin{equation}
    C_{\Gamma}(t)=\sum_{\vec x} \Big\langle [D_\Gamma Q](\vec x,t)~[D_\Gamma Q]^\dagger(\vec 0,0)  \Big\rangle~~.
\end{equation}
In this case the correlation function permits a spectral decomposition of the form \cite{Orginos:2005vr,Alexandrou:2005zn,Alexandrou:2006cq,Green:2010vc} 
\begin{equation}
    C_{\Gamma}(t)\sim \exp\left[-t\left(m_{D_{\Gamma}} + m_Q + \mathcal{O}(m_Q^{-1})\right)\right]~~,
\end{equation}
which gives a gauge-invariant probe for the spectrum of a given diquark channel through mass differences in which the mass of the static quark is exactly cancelled.

The first properties that we report on are diquark-diquark and diquark-quark mass differences involving diquark pairs with flavors $ud$, $\ell s~(\ell=u,d)$  and $ss'$. They can be accessed by taking the ratios of two diquark channel correlation functions, e.g. the "bad" ($\Gamma=\gamma_i$) and the good ($\Gamma=\gamma_5$), or a diquark and a static-light meson ($M_\Gamma=[\bar Q \Gamma q]$).

Given their fully non-perturbative origin and unique values in nature, 
the mass splittings can be viewed as fundamental characteristics of QCD \cite{Jaffe:2004ph}. Here, we present their calculation at a single lattice spacing. However, we expect discretisation effects only at the percent level given other calculations of the hadron spectrum on the same gauge configurations \cite{Padmanath:2019ybu,Namekawa:2013vu,Alexandrou:2017xwd,Hudspith:2017bbh}. By leveraging the large range of pion masses available to us we perform short, controlled chiral extrapolations to the physical value of the pion mass\footnote{We refer the reader to \cite{Francis:2021vrr} for further details.}. 
The results are summarised in Table 2, which show the mass differences calculated in our lattice study compared to their phenomenological counterparts. Overall we observe very good agreement with phenomenological 
expectations, and confirm the special
role of the attractive good diquark configuration.

\subsection{Diquark structure}

Going further, as validated by our
success in reproducing phenomenological expectations for the static limit spectroscopic splittings, we study the spatial correlations of the quarks embedded in the baryon with operator $B=[D_{\Gamma}Q]$. As before $\Gamma$
characterizes the diquark channel. 

\hspace{-4ex}
\begin{minipage}[t!]{0.49\textwidth}
\begin{center}
{\includegraphics[width=0.8\columnwidth]{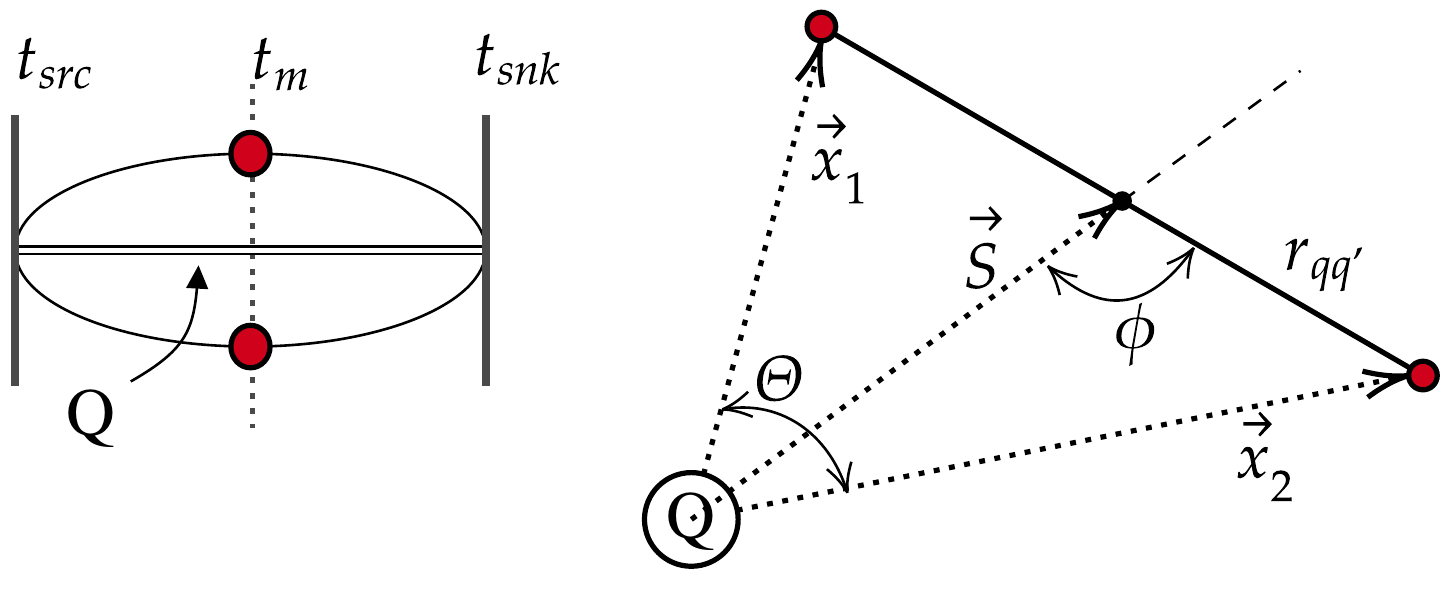}}
\end{center}
{Figure 1. {\it Sketch of the 
density correlators: 2D temporal
view (left) and current insertions, spatial view (right).}}
\end{minipage}
\vspace{2ex}

\noindent The quark density-density 
correlators 
\begin{align}
    C_{\Gamma}^{dd}(\vec x_1, \vec x_2, t)=
    \Big\langle B_\Gamma(\vec 0,2t)
    \rho(\vec x_1,t)\rho(\vec x_2,t)
    B_\Gamma^\dagger(\vec 0,0)  \Big\rangle
\end{align}
where $\rho(\vec x,t)=\bar{q}(\vec x,t)\gamma_0 q(\vec x,t)$, give a handle on the diquark's internal structure by studying their spatial properties. Their geometry is sketched in Figure 1, where we show a visualisation of the density-density correlators in temporal view (left) and in a spatial plane projection (right). 

The static quark is set at the origin, while the
light-quark source and sink points 
are located at $(\vec 0,t_{src})$ and 
$(\vec 0,t_{snk})$. We insert the currents at
$t_m=( t_{snk}+t_{src} )/2$ with $( t_{snk}-t_{src} )=16$. This maximises the projection onto the ground state in our simulation while keeping the noise manageable. We note that our calculations
average over all spatial translations, i.e. spatial positions of the quark sources and sinks. 
In the spatial view the relative positions 
of the static source and current
insertions $\vec{x}_1$, $\vec{x}_2$, 
can be understood in terms of
$\vec{r}_{qq^\prime}=\vec{x}_2-\vec{x}_1$,
$\vec{S}=(\vec{x}_1+\vec{x}_2)/2$, i.e.
the separation between the static source
and diquark midpoint, in addition to the angle $\phi$ between $\vec{r}_{qq^\prime}$
and $\vec{S}$.  
%
With this we define:
\begin{equation}
    \rho_2(r_{qq^\prime},S,\phi ;\Gamma ) \equiv 
    C^{dd}_\Gamma (\vec x_1, \vec x_2,t_m)\, .
\end{equation}
In this notation the distance from the static source to the 
closer of the two insertion points is 
minimized for $\phi =\pi$ and maximized for $\phi =\pi/2$ for a fixed $S$ and $r_{qq^\prime}$.
Note, the static quark could potentially disrupt the diquark correlation if they get too close. This disruption will therefore be largest for the angle $\phi =\pi$ 
and smallest for $\phi =\pi /2$.
In our study of $\rho_2$ we focus on these two limiting cases. In the case of smallest disruption, the distance 
$\vert \vec{x}_1\vert = 
\vert \vec{x}_2\vert\equiv R$ and the angle 
$\Theta$ between $\vec{x}_1$ and 
$\vec{x}_2$ may be used to characterise the correlations. Introducing a further shorthand we write in the following 
$\rho_2^\perp (R,\theta )\equiv 
\rho_2(r_{qq^\prime},S,\pi /2)$
and $\rho_2^\parallel(r_{qq^\prime},S)
\equiv \rho_2(r_{qq^\prime},S,\pi )$.

The correlator $\rho_2^\perp (R,\theta )$ is well suited to study possible quark-quark attraction effects. In case of attraction in a given diquark channel we expect an increase in $\rho_2^\perp (R,\theta )$ with decreasing $\Theta$ at fixed $R$. We show the results for all available diquark channels in Figure 2 (top) at a pion mass of $575$~MeV. 
Here the angular variable $\cos(\Theta)=-1\,(+1)$ implies the quarks are opposite (on top) of each other, i.e. $\Theta=180^\circ$ and $0^\circ$. We observe a clear increase exclusively in the good diquark channels $\Gamma=\gamma_5$ and $\gamma_5\gamma_0$. In all other channels we observe no indication of any such
attraction.

\hspace{-4ex}
 \begin{minipage}[t!]{0.49\textwidth}
\begin{center}
{\includegraphics[width=0.99\columnwidth]{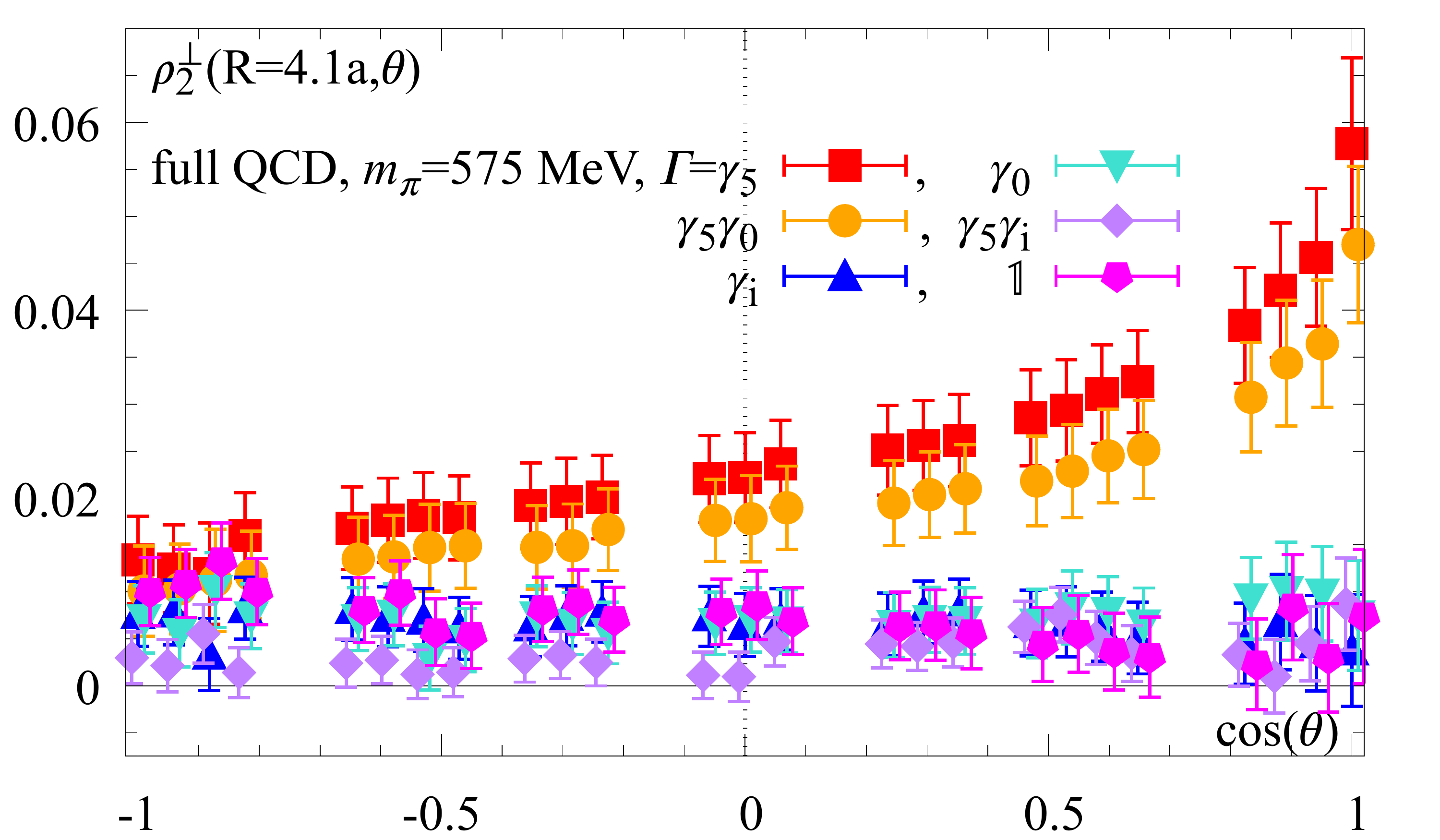}}
{\includegraphics[width=0.99\columnwidth]{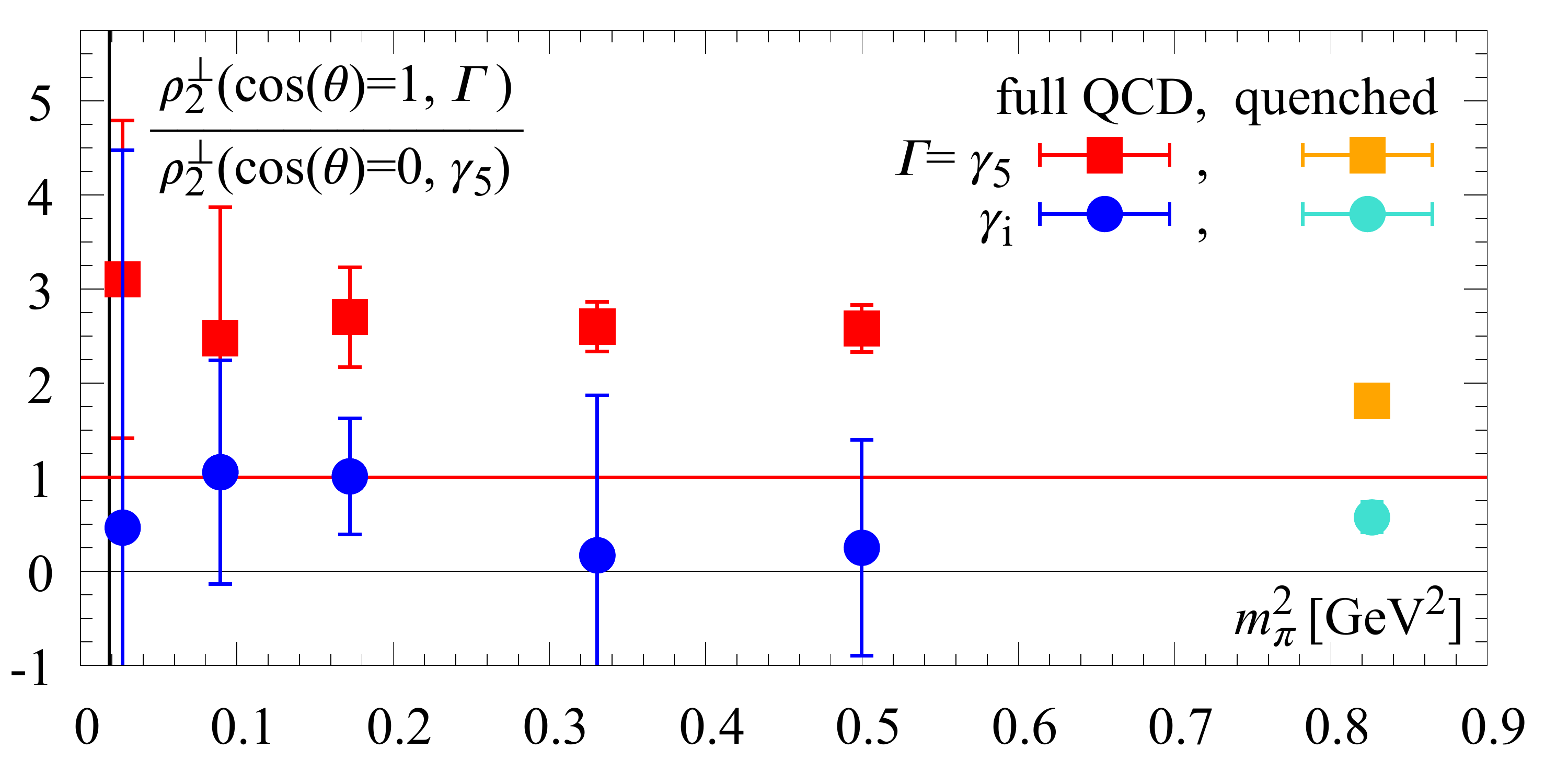}}
\end{center}
{Figure 2. {\it Diquark attraction. (Top) The density-density correlators
$\rho_2^\perp (R=4.1 a,\Theta ,\Gamma )$ versus 
$cos(\Theta )$ at $m_\pi=575~\rm{MeV}$. 
(Bottom) The ratio
$\rho_2^\perp (R,\Theta =0,\Gamma )/
\rho_2^\perp(R,\Theta =\pi /2, \Gamma =\gamma_5)$
versus $m_\pi^2$. Values above/below 1 for 
the red/blue points signal
attraction in the good diquark that is 
absent for the bad diquark. The vertical line
denotes physical $m_\pi$. 
}}
\label{fig:dens_panel}
 \end{minipage}
\vspace{2ex}

In Figure 2 (bottom) we study the quark mass dependence of this effect through the ratio between $\Theta=0^\circ$ and $90^\circ$: 
\begin{equation}
\frac{\rho_2^\perp(R,\Theta=0,\Gamma)}{\rho_2^\perp(R,\Theta=\pi/2,\gamma_5)}~.
\end{equation}
Focusing on the good, $\Gamma=\gamma_5$, and bad, $\Gamma=\gamma_i$, channels, we observe the good channel exhibits a significantly increased ratio for all masses available while it is small, consistent with zero, in the bad channel.
These observations establish the attractive interaction in the good diquark channel.

This picture of the good diquark can be further refined. Notice that the distance between the quarks in the probed diquark can be written as $r_{qq^\prime}=R\sqrt{2(1-\cos(\Theta ))}$. As such we can re-interpret our results as 
\begin{equation}
    \rho_2^\perp(R,r_{qq^\prime})\sim \exp(-r_{qq^\prime}/r_0)~~,
\end{equation}
where we defined the diquark size parameter $r_0$
through the scale of the exponential decay of the spatial correlation between the two quarks $q$ and $q^\prime$ constituting the diquark with $r_{qq^\prime}$. Our results for all available $R$ and $m_\pi$ are shown in Figure 3 (top). 
Note that we do not see evidence for a distortion through the static quark as long as $r_{qq^\prime} < R $. 
With the definition of the diquark size 
we perform a combined fit to all available $R$ at a given value of $m_\pi$.

\hspace{-4ex}
 \begin{minipage}[t!]{0.49\textwidth}
\begin{center}
{\includegraphics[width=0.99\columnwidth]{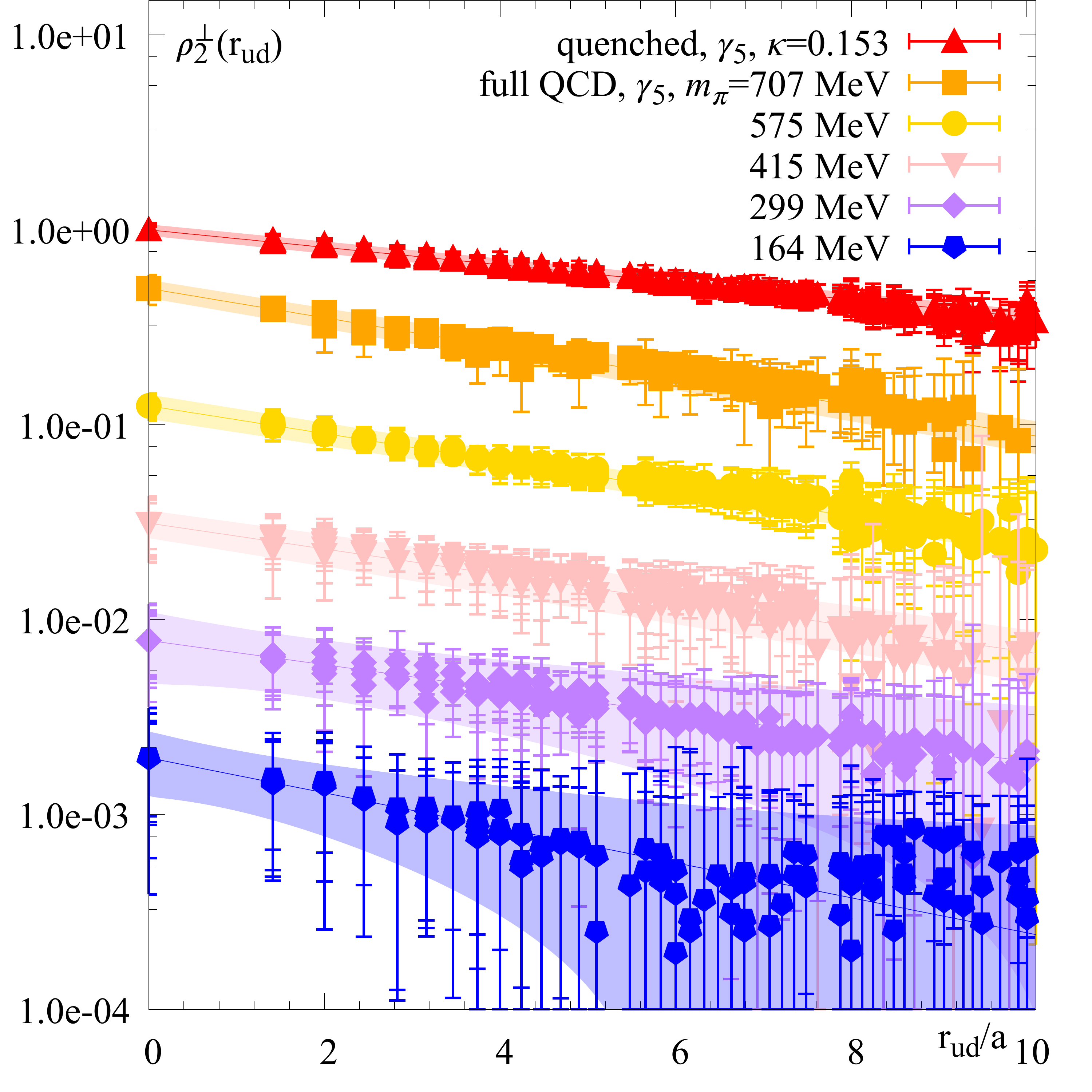}}
\vspace{-8pt}
{\includegraphics[width=0.99\columnwidth]{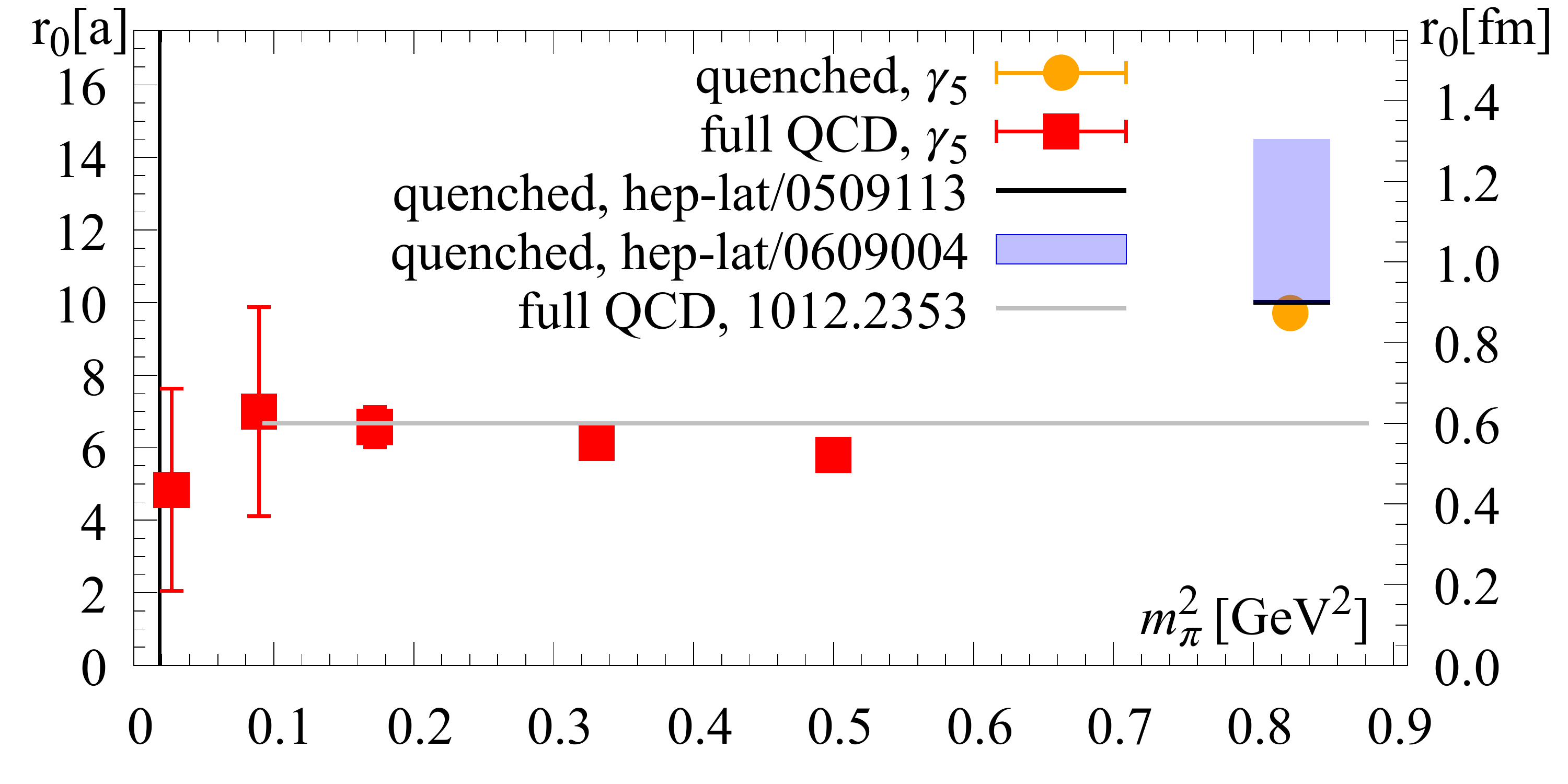}}
\end{center}
{Figure 3. {\it Good diquark size. (Top) 
Exponential decay with $r_{qq^\prime}$ 
of the density-density correlator
$\rho_2^\perp (R,\Theta )$. 
Each $m_\pi$ has its own color.
Data sets are normalised at 
$r_{qq^\prime}=0$ and offset vertically. 
Results for all available $R$ are shown 
together in one coloured set. Each coloured band
comes from the combined fit used to determine the diquark size $r_0(m_\pi^2)$. (Bottom) 
Resulting good diquark size $r_0$ versus
$m_\pi^2$, compared to results from the 
literature. The vertical line denotes physical
$m_\pi$. } }
\label{fig:rad_panel}
\end{minipage}
\vspace{2ex}

The results for $r_0(m_\pi^2)$ are displayed in Figure 3 (bottom), whereby we also compare with the results obtained in \cite{Alexandrou:2005zn,Alexandrou:2006cq,Green:2010vc}. Overall we observe very good agreement with these previous studies and significantly extend them.
Studying the decay of the spatial correlation between the quark-quark pair with distance we find the diameter of the diquark is $\sim 0.6$~fm. A similar value was found in determinations of the size of mesons and baryon using a similar method in \cite{Blossier:2016vgh}. As such, the good diquark is of hadronic size.

Finally, we further study the good diquarks by comparing analogue definitions of the diquark sizes separately in the relative radial (${r_0^\parallel}$, $\phi =\pi$) and tangential ($r_0^\perp$, $\phi =\pi /2$) orientations. This enables an estimation of the shape of the diquark and sheds light on possible polarisation effects through the static quark at the origin. 

\hspace{-4ex}
\begin{minipage}[t!]{0.49\textwidth}
\begin{center}
\includegraphics[width=0.99\columnwidth]{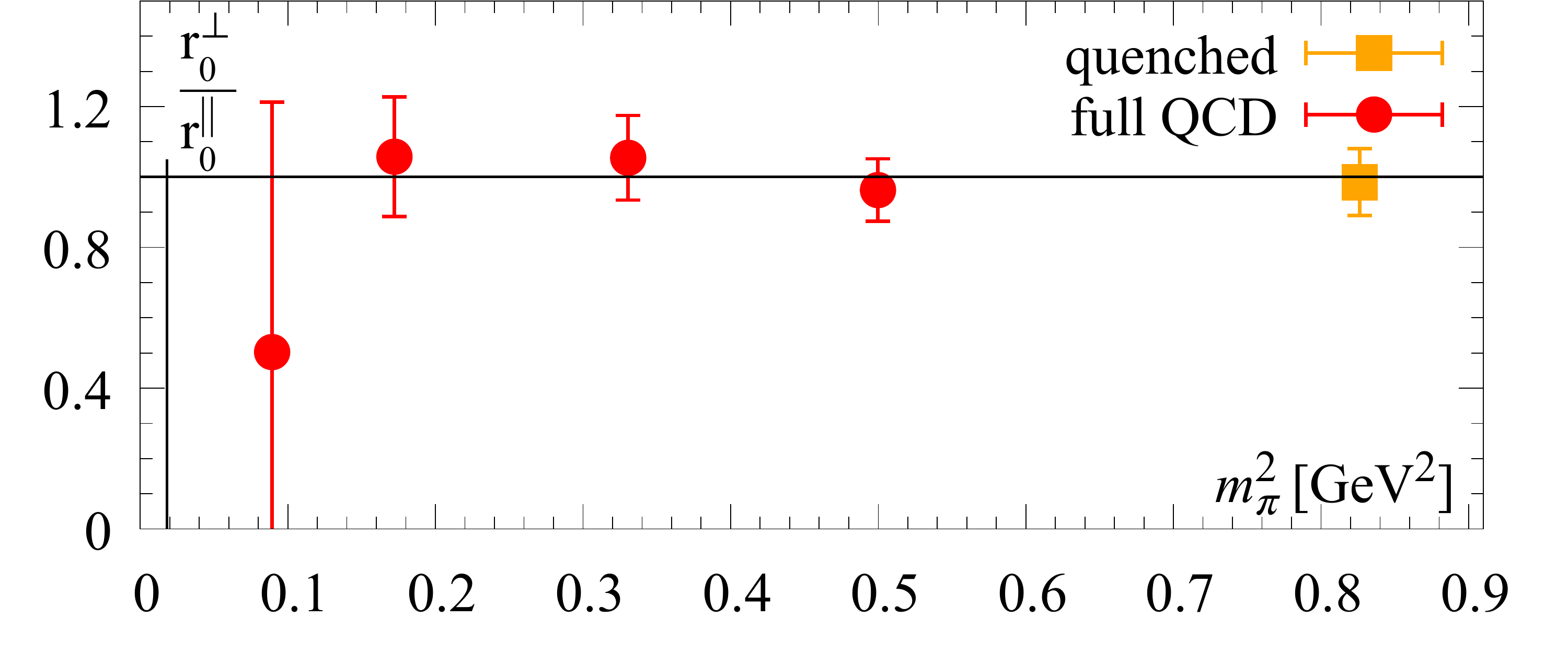}
\end{center}
{Figure 4. {\it Good diquark shape. The ratio 
$r_0^\perp/r_0^\parallel$ as
a function of $m_\pi$. 
The vertical line denotes physical 
$m_\pi$.}
}
\label{fig:obl_panel}
\end{minipage}
\vspace{2ex}

In particular the ratio $r_0^\perp/r_0^\parallel$
provides a measure of whether 
the diquarks are prolate, oblate, or 
spherical. The results are shown in Figure 4 and we find $r_0^\perp/r_0^\parallel(m_\pi^2)
\simeq 1$ within errors for all $m_\pi$. 
This indicates that the diquarks have a near-spherical shape
and that we do not observe polarisation effects due to the presence of the static quark.

\section{Discussion and Summary}

In this contribution, we presented results 
on both diquark spectroscopy and 
diquark structure, using ab initio lattice QCD simulations. By embedding the diquarks in baryons together with a single static quark we formed gauge-invariant probes of their properties. In particular we evaluated diquark-diquark and diquark-quark mass differences in which the static
spectator quark mass cancels out exactly. 
Comparing with phenomenological estimates we observe very good agreement for all available mass splittings. The splittings in particular confirm the special status of the good diquark, where we observe a relative mass difference of $198(4)$~MeV compared to the bad diquark after a short, controlled extrapolation to the physical pion mass point.

Going further we studied density-density correlations and found strong indication for quark-quark attraction in the good diquark channel. It should be stressed that this compact spatial correlation was observed only in the good diquark channel, once more providing clear, quantitative support for the good diquark picture. 
Defining the good diquark diameter through this spatial correlation we find it to roughly have the size $r_0\simeq \mathcal{O}(0,6)$~fm. This implies diquarks are similar in size to mesons and baryons \cite{Blossier:2016vgh}.
Finally we probed the shape of the good diquark by evaluating the ratio of its tangential and radial sizes. Our results imply an almost spherical shape, with no discernible
evidence for polarisation induced by the presence of the static spectator quark.


Further details and supporting studies that go beyond the scope of this contribution can found in \cite{Francis:2021vrr}. The results shown here were taken from this main reference.

\section*{Acknowledgements}

The authors acknowledge the support by the high performance computing resources Niagara maintained by SciNet and Compute Canada as well as HPC-QCD by CERN. RL and KM acknowledge the support
of grants from the Natural Sciences and 
Engineering Research Council of Canada.

\end{multicols}
\medline

\begin{multicols}{2}

\bibliographystyle{JHEP}
\bibliography{references}

\providecommand{\href}[2]{#2}\begingroup\raggedright\begin{thebibliography}{10}

\bibitem{Lichtenberg:1967zz}
D.B.~Lichtenberg and L.J.~Tassie, \emph{{Baryon Mass Splitting in a
  Boson-Fermion Model}},
  \href{https://doi.org/10.1103/PhysRev.155.1601}{\emph{Phys. Rev.} {\bfseries
  155} (1967) 1601}.

\bibitem{Jaffe:2004ph}
R.~Jaffe, \emph{{Exotica}},
  \href{https://doi.org/10.1016/j.physrep.2004.11.005}{\emph{Phys. Rept.}
  {\bfseries 409} (2005) 1}
  [\href{https://arxiv.org/abs/hep-ph/0409065}{{\ttfamily hep-ph/0409065}}].

\bibitem{Francis:2016hui}
A.~Francis, R.J.~Hudspith, R.~Lewis and K.~Maltman, \emph{{Lattice Prediction
  for Deeply Bound Doubly Heavy Tetraquarks}},
  \href{https://doi.org/10.1103/PhysRevLett.118.142001}{\emph{Phys. Rev. Lett.}
  {\bfseries 118} (2017) 142001}
  [\href{https://arxiv.org/abs/1607.05214}{{\ttfamily 1607.05214}}].

\bibitem{Karliner:2017qjm}
M.~Karliner and J.L.~Rosner, \emph{{Discovery of doubly-charmed $\Xi_{cc}$
  baryon implies a stable ($b b \bar{u} \bar{d}$) tetraquark}},
  \href{https://doi.org/10.1103/PhysRevLett.119.202001}{\emph{Phys. Rev. Lett.}
  {\bfseries 119} (2017) 202001}
  [\href{https://arxiv.org/abs/1707.07666}{{\ttfamily 1707.07666}}].

\bibitem{Eichten:2017ffp}
E.J.~Eichten and C.~Quigg, \emph{{Heavy-quark symmetry implies stable heavy
  tetraquark mesons $Q_iQ_j \bar q_k \bar q_l$}},
  \href{https://doi.org/10.1103/PhysRevLett.119.202002}{\emph{Phys. Rev. Lett.}
  {\bfseries 119} (2017) 202002}
  [\href{https://arxiv.org/abs/1707.09575}{{\ttfamily 1707.09575}}].

\bibitem{Czarnecki:2017vco}
A.~Czarnecki, B.~Leng and M.B.~Voloshin, \emph{{Stability of tetrons}},
  \href{https://doi.org/10.1016/j.physletb.2018.01.034}{\emph{Phys. Lett. B}
  {\bfseries 778} (2018) 233}
  [\href{https://arxiv.org/abs/1708.04594}{{\ttfamily 1708.04594}}].

\bibitem{Francis:2018jyb}
A.~Francis, R.J.~Hudspith, R.~Lewis and K.~Maltman, \emph{{Evidence for
  charm-bottom tetraquarks and the mass dependence of heavy-light tetraquark
  states from lattice QCD}},
  \href{https://doi.org/10.1103/PhysRevD.99.054505}{\emph{Phys. Rev. D}
  {\bfseries 99} (2019) 054505}
  [\href{https://arxiv.org/abs/1810.10550}{{\ttfamily 1810.10550}}].

\bibitem{Hudspith:2020tdf}
R.J.~Hudspith, B.~Colquhoun, A.~Francis, R.~Lewis and K.~Maltman, \emph{{A
  lattice investigation of exotic tetraquark channels}},
  \href{https://doi.org/10.1103/PhysRevD.102.114506}{\emph{Phys. Rev. D}
  {\bfseries 102} (2020) 114506}
  [\href{https://arxiv.org/abs/2006.14294}{{\ttfamily 2006.14294}}].

\bibitem{Junnarkar:2018twb}
P.~Junnarkar, N.~Mathur and M.~Padmanath, \emph{{Study of doubly heavy
  tetraquarks in Lattice QCD}},
  \href{https://doi.org/10.1103/PhysRevD.99.034507}{\emph{Phys. Rev. D}
  {\bfseries 99} (2019) 034507}
  [\href{https://arxiv.org/abs/1810.12285}{{\ttfamily 1810.12285}}].

\bibitem{Bicudo:2012qt}
{\scshape European Twisted Mass} collaboration, \emph{{Lattice QCD signal for a
  bottom-bottom tetraquark}},
  \href{https://doi.org/10.1103/PhysRevD.87.114511}{\emph{Phys. Rev. D}
  {\bfseries 87} (2013) 114511}
  [\href{https://arxiv.org/abs/1209.6274}{{\ttfamily 1209.6274}}].

\bibitem{Bicudo:2015kna}
P.~Bicudo, K.~Cichy, A.~Peters and M.~Wagner, \emph{{BB interactions with
  static bottom quarks from Lattice QCD}},
  \href{https://doi.org/10.1103/PhysRevD.93.034501}{\emph{Phys. Rev. D}
  {\bfseries 93} (2016) 034501}
  [\href{https://arxiv.org/abs/1510.03441}{{\ttfamily 1510.03441}}].

\bibitem{Bicudo:2015vta}
P.~Bicudo, K.~Cichy, A.~Peters, B.~Wagenbach and M.~Wagner, \emph{{Evidence for
  the existence of $u d \bar{b} \bar{b}$ and the non-existence of $s s \bar{b}
  \bar{b}$ and $c c \bar{b} \bar{b}$ tetraquarks from lattice QCD}},
  \href{https://doi.org/10.1103/PhysRevD.92.014507}{\emph{Phys. Rev. D}
  {\bfseries 92} (2015) 014507}
  [\href{https://arxiv.org/abs/1505.00613}{{\ttfamily 1505.00613}}].

\bibitem{Bicudo:2016ooe}
P.~Bicudo, J.~Scheunert and M.~Wagner, \emph{{Including heavy spin effects in
  the prediction of a $\bar{b} \bar{b} u d$ tetraquark with lattice QCD
  potentials}}, \href{https://doi.org/10.1103/PhysRevD.95.034502}{\emph{Phys.
  Rev. D} {\bfseries 95} (2017) 034502}
  [\href{https://arxiv.org/abs/1612.02758}{{\ttfamily 1612.02758}}].

\bibitem{Bicudo:2021qxj}
P.~Bicudo, A.~Peters, S.~Velten and M.~Wagner, \emph{{Importance of meson-meson
  and of diquark-antidiquark creation operators for a $\bar{b} \bar{b} u d$
  tetraquark}},  \href{https://arxiv.org/abs/2101.00723}{{\ttfamily
  2101.00723}}.

\bibitem{Leskovec:2019ioa}
L.~Leskovec, S.~Meinel, M.~Pflaumer and M.~Wagner, \emph{{Lattice QCD
  investigation of a doubly-bottom $\bar{b} \bar{b} u d$ tetraquark with
  quantum numbers $I(J^P) = 0(1^+)$}},
  \href{https://doi.org/10.1103/PhysRevD.100.014503}{\emph{Phys. Rev. D}
  {\bfseries 100} (2019) 014503}
  [\href{https://arxiv.org/abs/1904.04197}{{\ttfamily 1904.04197}}].

\bibitem{Francis:2021vrr}
A.~Francis, P.~de~Forcrand, R.~Lewis and K.~Maltman, \emph{{Diquark properties
  from full QCD lattice simulations}},
  \href{https://arxiv.org/abs/2106.09080}{{\ttfamily 2106.09080}}.

\bibitem{Aoki:2008sm}
{\scshape PACS-CS} collaboration, \emph{{2+1 Flavor Lattice QCD toward the
  Physical Point}},
  \href{https://doi.org/10.1103/PhysRevD.79.034503}{\emph{Phys. Rev. D}
  {\bfseries 79} (2009) 034503}
  [\href{https://arxiv.org/abs/0807.1661}{{\ttfamily 0807.1661}}].

\bibitem{Namekawa:2013vu}
{\scshape PACS-CS} collaboration, \emph{{Charmed baryons at the physical point
  in 2+1 flavor lattice QCD}},
  \href{https://doi.org/10.1103/PhysRevD.87.094512}{\emph{Phys. Rev. D}
  {\bfseries 87} (2013) 094512}
  [\href{https://arxiv.org/abs/1301.4743}{{\ttfamily 1301.4743}}].

\bibitem{JLDG}
JLDG, \emph{Ensembles available from \url{https://www.jldg.org}}, .

\bibitem{Alexandrou:2005zn}
C.~Alexandrou, P.~de~Forcrand and B.~Lucini, \emph{{Searching for diquarks in
  hadrons}}, \href{https://doi.org/10.22323/1.020.0053}{\emph{PoS} {\bfseries
  LAT2005} (2006) 053} [\href{https://arxiv.org/abs/hep-lat/0509113}{{\ttfamily
  hep-lat/0509113}}].

\bibitem{Alexandrou:2006cq}
C.~Alexandrou, P.~de~Forcrand and B.~Lucini, \emph{{Evidence for diquarks in
  lattice QCD}},
  \href{https://doi.org/10.1103/PhysRevLett.97.222002}{\emph{Phys. Rev. Lett.}
  {\bfseries 97} (2006) 222002}
  [\href{https://arxiv.org/abs/hep-lat/0609004}{{\ttfamily hep-lat/0609004}}].

\bibitem{Hess:1998sd}
M.~Hess, F.~Karsch, E.~Laermann and I.~Wetzorke, \emph{{Diquark masses from
  lattice QCD}}, \href{https://doi.org/10.1103/PhysRevD.58.111502}{\emph{Phys.
  Rev.} {\bfseries D58} (1998) 111502}
  [\href{https://arxiv.org/abs/hep-lat/9804023}{{\ttfamily hep-lat/9804023}}].

\bibitem{Bi:2015ifa}
Y.~Bi, H.~Cai, Y.~Chen, M.~Gong, Z.~Liu, H.-X.~Qiao et~al., \emph{{Diquark mass
  differences from unquenched lattice QCD}},
  \href{https://doi.org/10.1088/1674-1137/40/7/073106}{\emph{Chin. Phys. C}
  {\bfseries 40} (2016) 073106}
  [\href{https://arxiv.org/abs/1510.07354}{{\ttfamily 1510.07354}}].

\bibitem{Babich:2007ah}
R.~Babich, N.~Garron, C.~Hoelbling, J.~Howard, L.~Lellouch and C.~Rebbi,
  \emph{{Diquark correlations in baryons on the lattice with overlap quarks}},
  \href{https://doi.org/10.1103/PhysRevD.76.074021}{\emph{Phys. Rev.}
  {\bfseries D76} (2007) 074021}
  [\href{https://arxiv.org/abs/hep-lat/0701023}{{\ttfamily hep-lat/0701023}}].

\bibitem{Teo:1992zu}
K.B.~Teo and J.W.~Negele, \emph{{The Definition and lattice measurement of
  hadron wave functions}},
  \href{https://doi.org/10.1016/0920-5632(94)90399-9}{\emph{Nucl. Phys. Proc.
  Suppl.} {\bfseries 34} (1994) 390}.

\bibitem{Negele:2000uk}
J.W.~Negele, \emph{{Hadron structure in lattice QCD: Exploring the gluon wave
  functional}},  in \emph{{Excited nucleons and hadronic structure.
  Proceedings, Conference, NSTAR 2000, Newport News, USA, February 16-19,
  2000}}, pp.~368--377, 2000
  [\href{https://arxiv.org/abs/hep-lat/0007026}{{\ttfamily hep-lat/0007026}}].

\bibitem{Alexandrou:2002nn}
C.~Alexandrou, P.~de~Forcrand and A.~Tsapalis, \emph{{Probing hadron wave
  functions in lattice QCD}},
  \href{https://doi.org/10.1103/PhysRevD.66.094503}{\emph{Phys. Rev.}
  {\bfseries D66} (2002) 094503}
  [\href{https://arxiv.org/abs/hep-lat/0206026}{{\ttfamily hep-lat/0206026}}].

\bibitem{Zyla:2020zbs}
{\scshape Particle Data Group} collaboration, \emph{{Review of Particle
  Physics}}, \href{https://doi.org/10.1093/ptep/ptaa104}{\emph{PTEP} {\bfseries
  2020} (2020) 083C01}.

\bibitem{Orginos:2005vr}
K.~Orginos, \emph{{Diquark properties from lattice QCD}},
  \href{https://doi.org/10.22323/1.020.0054}{\emph{PoS} {\bfseries LAT2005}
  (2006) 054} [\href{https://arxiv.org/abs/hep-lat/0510082}{{\ttfamily
  hep-lat/0510082}}].

\bibitem{Green:2010vc}
J.~Green, J.~Negele, M.~Engelhardt and P.~Varilly, \emph{{Spatial diquark
  correlations in a hadron}},
  \href{https://doi.org/10.22323/1.105.0140}{\emph{PoS} {\bfseries LATTICE2010}
  (2010) 140} [\href{https://arxiv.org/abs/1012.2353}{{\ttfamily 1012.2353}}].

\bibitem{Padmanath:2019ybu}
M.~Padmanath, \emph{{Heavy baryon spectroscopy from lattice QCD}},
  \href{https://arxiv.org/abs/1905.10168}{{\ttfamily 1905.10168}}.

\bibitem{Alexandrou:2017xwd}
C.~Alexandrou and C.~Kallidonis, \emph{{Low-lying baryon masses using $N_f=2$
  twisted mass clover-improved fermions directly at the physical pion mass}},
  \href{https://doi.org/10.1103/PhysRevD.96.034511}{\emph{Phys. Rev. D}
  {\bfseries 96} (2017) 034511}
  [\href{https://arxiv.org/abs/1704.02647}{{\ttfamily 1704.02647}}].

\bibitem{Hudspith:2017bbh}
R.J.~Hudspith, A.~Francis, R.~Lewis and K.~Maltman, \emph{{Heavy and light
  spectroscopy near the physical point, Part I: Charm and bottom baryons}},
  \href{https://doi.org/10.22323/1.256.0133}{\emph{PoS} {\bfseries LATTICE2016}
  (2017) 133}.

\bibitem{Blossier:2016vgh}
B.~Blossier and A.~G\'erardin, \emph{{Density distributions in the $B$ meson}},
  \href{https://doi.org/10.1103/PhysRevD.94.074504}{\emph{Phys. Rev. D}
  {\bfseries 94} (2016) 074504}
  [\href{https://arxiv.org/abs/1604.02891}{{\ttfamily 1604.02891}}].

\end{thebibliography}\endgroup

%

\end{multicols}

\end{document}